\documentstyle[11pt,IAU207_pasp,psfig,twoside]{article}
\def\edcomment#1{\iffalse\marginpar{\raggedright\sl#1\/}\else\relax\fi}
\reversemarginpar

\begin{document}
\title{Ultra-compact dwarf galaxies:
a new class of compact stellar system discovered in the Fornax 
Cluster}
\author{Michael Drinkwater}
\affil{School of Physics, University of Melbourne, Victoria 3010, Australia}
\author{Kenji Bekki, Warrick Couch}
\affil{School of Physics, Univ.\ of New South Wales, Sydney 2052, Australia}
\author{Steve Phillipps}
\affil{Department of Physics, University of Bristol,
Bristol BS8 1TL, UK}
\author{Bryn Jones}
\affil{School of Physics, University of Nottingham, Nottingham NG7 2RD, UK}
\author{Michael Gregg}
\affil{Department of Physics, University of California, Davis, CA
95616, USA}

\begin{abstract}We have used the 2dF spectrograph
on the Anglo-Australian Telescope to obtain a complete
spectroscopic sample of all objects in the magnitude range,
$16.5<b_J<19.8$, regardless of morphology, in an area centred
on the Fornax Cluster of galaxies. Among the unresolved targets
are five objects which are members
of the Fornax Cluster. They are extremely compact
stellar systems with scale lengths less than 40 parsecs.  These
ultra-compact dwarfs are unlike any known type
of stellar system, being more compact and significantly less luminous
than other compact dwarf galaxies, yet much brighter than any globular
cluster.
\end{abstract}

\begin{figure}
\centerline{\vbox{
\psfig{figure=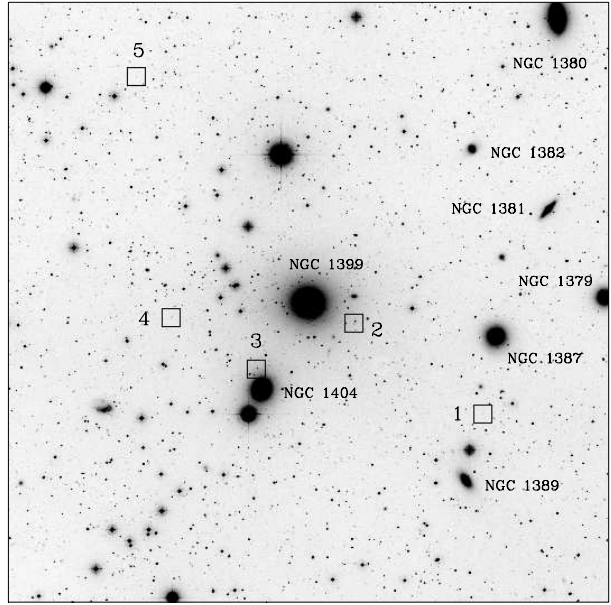,width=8cm}
}}
\caption{R-band UKST image
of the central 1 degree region of the Fornax Cluster. The positions
of the UCDs are shown by squares.}
\end{figure}

\section{The Fornax Cluster Spectroscopic Survey}

Our 2dF Fornax Cluster Spectroscopic Survey (FCSS; Drinkwater et al.\
2000a) was designed to make the most complete census possible of
low-luminosity galaxies in the Fornax Cluster. Having shown
(Drinkwater \& Gregg 1998) that compact, high surface brightness dwarf
cluster galaxies were missed in previous work, we took the unusual
step of observing {\em all} objects in each 2dF field, both resolved
(``galaxies'') and unresolved (``stars''). In this way we avoided any
morphological bias as to what a cluster galaxy should look
like. Observing the ``stars'' in each field greatly increases the
number of targets: over 50\% of objects with $16.5<b_J<19.8$ are
stars. However the
flexibility of 2dF means this only led to a small increase in
total observing time as we scheduled the stars at times when we could
not usefully observe galaxies, such as twilight or
through cloud.

\begin{figure}
\centerline{\hbox{
\psfig{figure=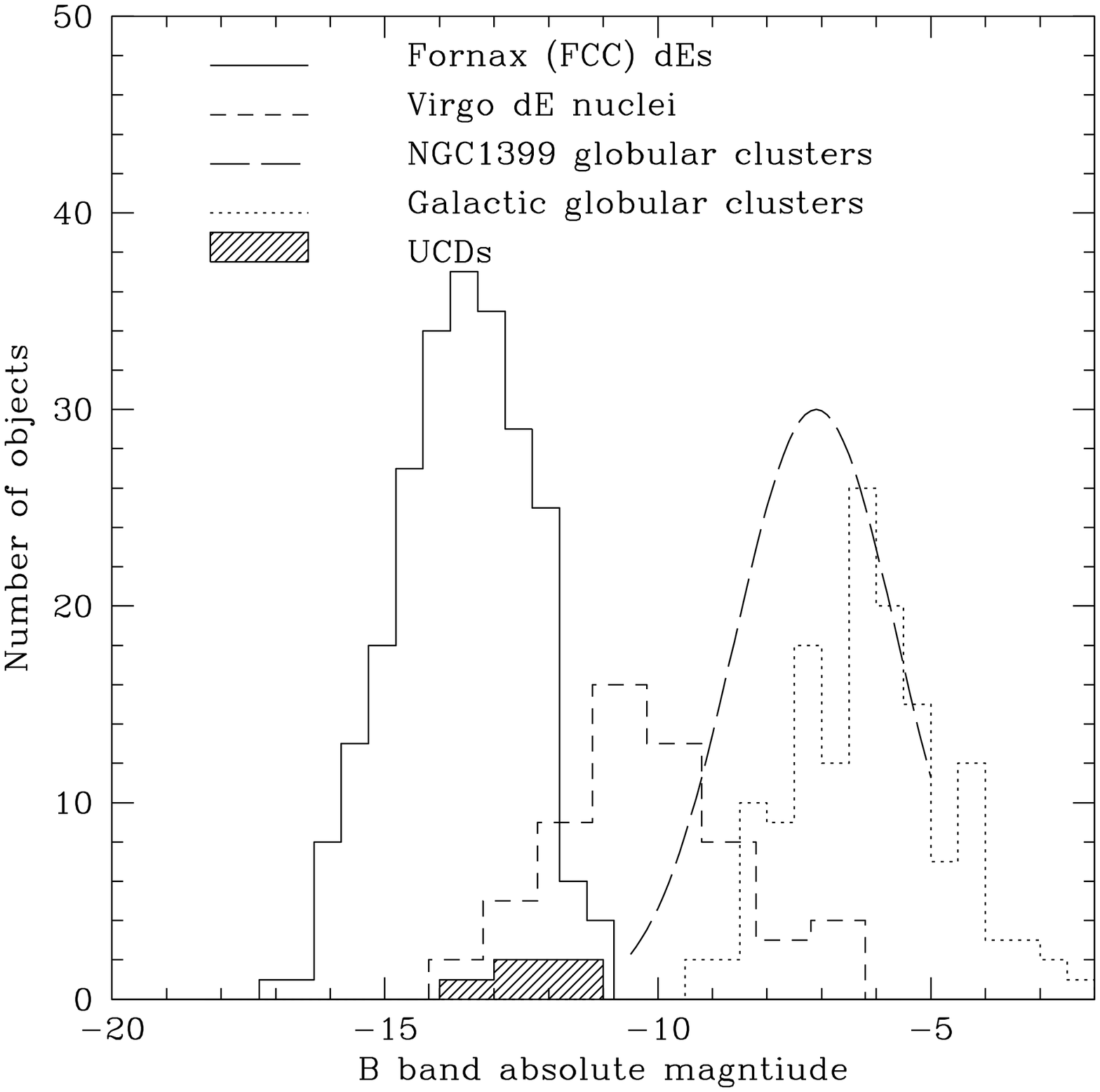,width=7cm}
\psfig{figure=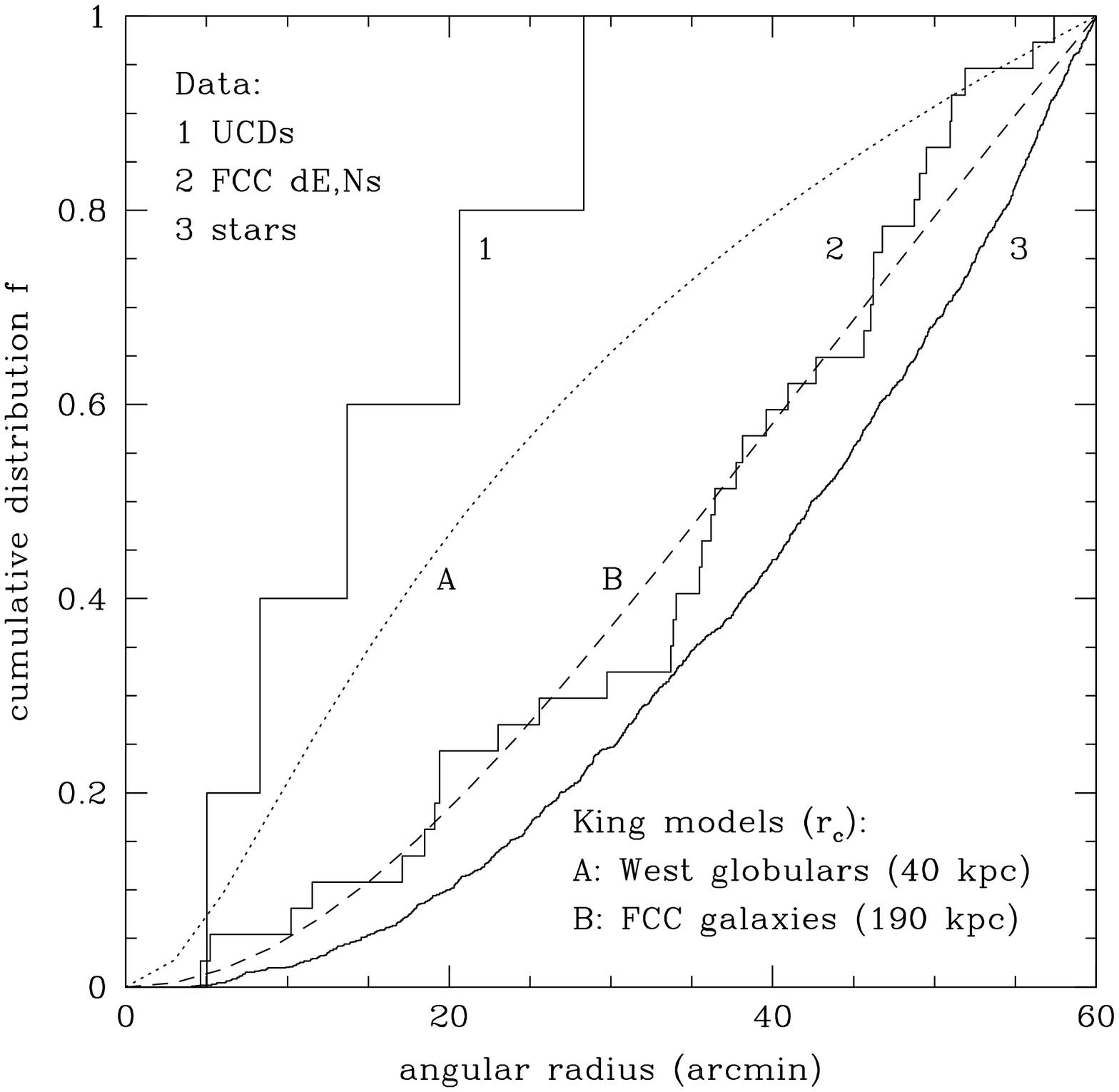,width=7cm}
}}
\caption{Left: Absolute magnitude distribution of UCDs
compared to Fornax Cluster dEs (FCC, Ferguson 1989), the nuclei
of Virgo Cluster dE,Ns (Binggeli \& Cameron 1991), a fit
to globular clusters around NGC 1399 (Bridges et al.\ 1991) and
Galactic globular clusters (Harris 1996). Right: Cumulative radial
distribution of the new compact objects compared to the predicted
distribution for intra-cluster globular clusters (West et al.\ 1995)
and the profile fit to the distribution of all FCC members. Also shown
is the distribution of all nucleated dwarfs in the FCC and the
stars observed in our survey.
\label{fig-props}}
\end{figure}

\section{Properties of Ultra-compact Dwarfs}

The highlight of our results to date has been the discovery of a new
population of Fornax Cluster dwarf galaxies so compact they were
previously mistaken for Galactic stars (Drinkwater et al.\ 2000b). The
locations of these objects in the cluster is shown in Fig.~1.  
The 2dF discovery spectra of these compact objects show them to be old
stellar systems---no Balmer lines are detected---but we await higher
resolution spectra for more detailed classifications. These
``ultra-compact dwarf'' (UCD) galaxies are unlike any known type of
stellar system. They are smaller and more concentrated than any known
dwarf galaxy, but are 2-3 magnitudes more luminous than the largest
Galactic globular clusters. Their luminosities are compared with other
stellar systems in Fig.~2(a). The only known objects they resemble
both in luminosity and morphology are the nuclei of nucleated dwarf
elliptical galaxies, but without any surrounding low surface
brightness envelope. The radial distribution of the UCDs given in
Fig.~2(b) shows that they are concentrated to the centre of the
cluster, much more than the general population of cluster galaxies or
even the nucleated dwarfs. The only distribution on the Figure that
resembles that of the UCDs is that proposed for globular clusters
formed in situ in the intra-cluster medium as suggested by West et
al.\ (1995).

\begin{figure}
\centerline{\vbox{
\psfig{figure=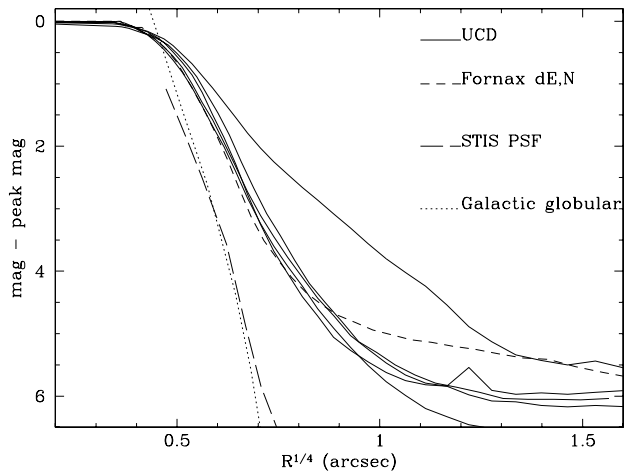,width=9cm}
}}
\caption{Radial profiles of the UCDs measured with STIS compared to
the centre of a nucleated dwarf elliptical galaxy in the Fornax
cluster measured with the same setup. The profiles are all normalised
to their peak intensity. Also shown are the instrumental PSF from the
STIS manual and the profile of a galactic globular cluster scaled to
the distance of Fornax but not convolved with the PSF.}
\end{figure}

We have obtained high-resolution images of the UCDs with the STIS
instrument on the {\em
Hubble Space Telescope} (HST) to measure their sizes.  From a
preliminary analysis of our HST data we have calculated
radial profiles of the UCDs shown in Fig.~3. Four of the UCDs and a
dE,N nucleus we observed for comparison all have very similar
profiles: they are resolved by STIS with FWHM=11--16 pc and de Vaucouleurs
effective radii of $R_e\approx10$ pc. One of the UCDs (number 3; see
Fig.~1) is larger than the rest (FWHM=17 pc, $R_e\approx300$
pc). The UCDs are all significantly larger than globular clusters, but
they do have similar physical sizes and profiles to the dwarf
elliptical nucleus we observed (except for number 3).

\begin{figure}
\centerline{\vbox{
\psfig{figure=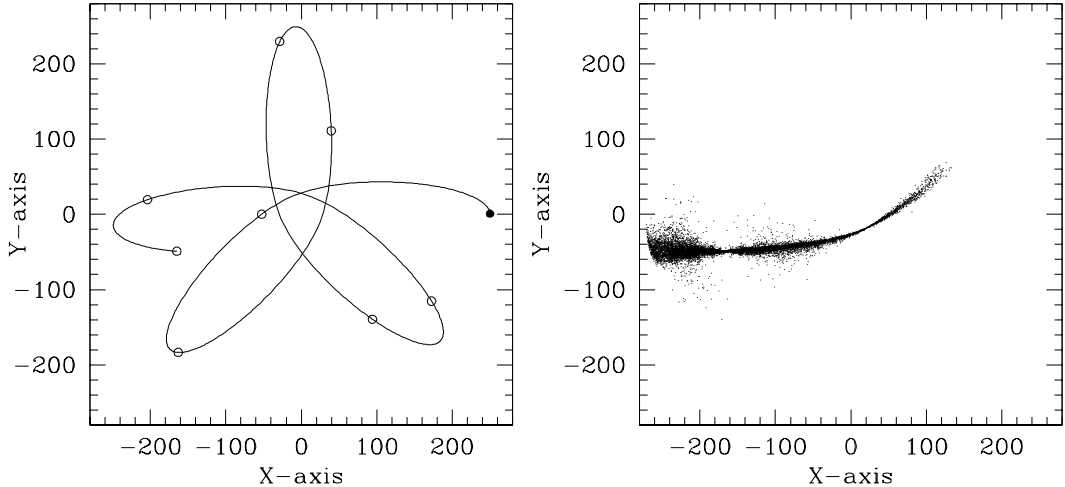,width=14cm}
}}
\caption{Numerical simulation of the formation of a UCD by tidal
stripping of a nucleated dwarf elliptical in an orbit passing close to
the central cluster galaxy NGC~1399 (see Bekki et al.\ 2001 for
details).  Left: The orbit of the simulated dwarf around NGC~1399 (at
the origin; the scale is in units of 0.8 kpc, so the frame is 450 kpc
across).  The filled circle represents the initial position of the
dwarf and the orbital evolution is indicated by open circles
indicating successive time intervals of 4.7 $\times$ $10^8$ yr.
Right: the final mass distribution of the dwarf galaxy at a time of
3.8 Gyr.  Owing to the strong tidal field of NGC~1399, the dwarf is
greatly stretched and most of the outer stellar component of the dwarf
is tidally stripped away. The core of the dwarf, however, remains
intact.}
\end{figure}

\section{What are Ultra-compact Dwarf galaxies?}

We are considering three hypotheses to explain the nature and
origin of the UCDs. First they may simply be very
massive star clusters, such as the intra-cluster
globular clusters proposed by West et al.\ (1995), although their
luminosities remain very high, well beyond those of galactic globular
clusters. Perhaps we are seeing the very high-luminosity tail of the
globular cluster population associated with NGC~1399.

Our second hypothesis is that these are the stripped nuclei of dwarf
galaxies.  Previous studies of the very large globular cluster
population associated with central cluster galaxies like NCG~1399 in
the Fornax Cluster have invoked tidal stripping of dwarf galaxies to
explain the globular cluster populations. This process was simulated
in some detail by Bassino et al.\ (1994) who predicted the formation
of even larger remnants that might resemble the UCDs. We have
performed numerical simulations of tidal stripping of nucleated dwarf
galaxies in close orbits around the central galaxy of the cluster,
NGC~1399 (Bekki, Couch \& Drinkwater 2001). The simulations show that
this process is a feasible mechanism for the formation of UCDs. The
result of one simulation is shown in Fig.~4.  The simulations show
that dwarfs in close orbits around NGC~1399 can have their outer
stellar envelopes removed in a few orbits (a few Gyr), leaving just
the nuclei which we have shown are of similar size to the UCDs.

Finally they may represent an entirely new type of compact galaxy. In
this case they mmight be miniature compact ellipticals (like M32). The
main test we plan of these different hypotheses is to measure the
internal dynamics of the UCDs using high-resolution VLT spectra to
measure their velocity dispersions. We will combine these with our HST
measurements of their sizes to make estimates of their virial masses
and hence their mass-to-light ratios (M/L). We would expect star
clusters to have low M/L values, whereas compact galaxies would have
higher M/L values, if we assume that they have some dark matter
content---there is still no clear prediction on the existence of dark
matter halos at these low mass limits.

\acknowledgements

This project has only been possible thanks to the superb performance of
the 2dF facility, for which we wish to thank the many staff of the
Anglo-Australian Observatory involved in its support. We also
acknowledge the assistance of our colleagues in the FCSS team. Their
names can be found on our web site at
http://astro.ph.unimelb.edu.au/data/
where we also provide public access to the data from our first field.

\end{document}